\documentclass[reprint, aps, prl, floatfix,superscriptaddress]{revtex4-2}
\usepackage{latexsym}
\usepackage{graphicx}    
\usepackage{dcolumn}    
\usepackage{bm}             
\usepackage{amssymb}   
\usepackage{amsmath}    
\usepackage{amsthm}
\usepackage{mathrsfs}    
\usepackage{commath}   
\usepackage{subfigure}    
\usepackage{braket}
\usepackage{ulem}
\usepackage{comment}
\usepackage{color}
\usepackage{siunitx}
\usepackage[colorlinks=true, allcolors=blue]{hyperref}
\usepackage{hyphenat}
\usepackage{footnote}
\usepackage{notes2bib}
\usepackage{xcolor}

\newcommand{\beq}{\begin{equation}}
\newcommand{\eeq}{\end{equation}}

\begin{document}

\title{Correlated noise can be beneficial to quantum transducers}

\author{Yu-Bo Hou}
\altaffiliation{These authors contributed equally to this work.}
\author{Xiaoan Ai}
\altaffiliation{These authors contributed equally to this work.}
\author{Ruizhe You}
\author{Changchun Zhong}
\email{zhong.changchun@xjtu.edu.cn}

\affiliation{Department of Physics, Xi'an Jiaotong University, Xi'an, Shanxi 710049, China}

\begin{abstract}
Quantum systems are inherently susceptible to noise—a notorious factor that induces decoherence and limits the performance of quantum applications. To mitigate its detrimental effects, various techniques have been developed, including cryogenic cooling, bath engineering, and quantum error correction. In this letter, we demonstrate that by exploiting noise correlations in coupled quantum systems, the overall impact of noise can be significantly suppressed. Specifically, in a piezo-optomechanical-based microwave-optical quantum transducer, correlations between the noise affecting the acoustic and electrical modes can lead to substantial noise reduction, thereby enhancing the performance of quantum transduction. This reduction is primarily governed by the phase of the piezo-mechanical coupling and is also influenced by system parameters such as the coupling ratio and mode cooperativity. Since these parameters simultaneously affect the signal transmissivity, they must be optimized to achieve the transducer's optimal performance. Our work provides a systematic framework for this optimization, offering a guidance for practical designs.  
\end{abstract}

\maketitle

\textit{Introduction}---Cooling a quantum system---such as a hybrid quantum transducer with coupled bosonic modes \cite{Andrews2014,qt1,qt2,qt3,qt4,zhong2022prapp,mirhosseini2020,Javerzac-Galy2016,Hisatomi2016}---close to its ground state is crucial for its applications in the quantum regime. At finite temperature, the bosonic modes may exhibit high thermal occupation, which would induce detrimental decoherence to the system, thereby hinder the execution of any quantum tasks. Cryogenic condition is typically achieved using dilute refrigerators capable of reaching temperatures of several millikelvin (or even microkelvin), albeit with limited cooling power \cite{radebaugh1971}. However, to have a better power-handling capability, these refrigerators generally operate more efficiently at slightly higher temperatures \cite{zu2022}. In addition, to maintain a strong bosonic mode coupling strength or to have an effective quantum operation, the slightly hot environment is sometimes unavoidable since controlled laser pumps have to be used which cause laser heating \cite{han2020,wei2021,zhong2020prl}. This indicates that the bosonic modes would have to frequently operate in a non-negligible thermal environment in practical situations. 

Thus, it is often necessary for the system to maintain low thermal excitation of its modes while residing in a higher-temperature environment. Radiative cooling is one technique to achieve this, and has been discussed in various contexts, including the setting of superconducting resonator \cite{mingrui2020}, cavity piezo-mechanics \cite{han2020,meesala2024prx}, and spin ensembles \cite{albanese2020,mark2021}. By coupling the system to an extra cold reservoir, the thermal occupation of specific modes can be radiatively cooled near their ground state, even when its surrounding remains relatively hot \cite{mingrui2020}. In this letter, we propose a new scheme that go beyond conventional radiative cooling by further suppressing thermal noise through the exploitation of noise correlations. Specifically, for the quantum transducers, the coupled bosonic modes will inevitably interact their noisy environment, which are potentially generated by similar physical process, thus exhibiting certain correlations in the noise. To the best of our knowledge, we propose a protocol, for the first time of its kind, that by properly tuning the mode coupling phase, the noise correlation can contribute to the reduction of overall thermal noise in the quantum transducer, thereby enhancing the signal conversion capability.

\textit{Quantum transduction model}---Quantum transducer aims to convert quantum information encoded in microwave ($\omega_\text{e}$) and optical ($\omega_\text{o}$) frequencies. Due to the large energy difference, the conversion is generally realized by nonlinear couplings between microwave and optical bosonic modes, e.g., the scattering pressure induced optomechanical coupling \cite{aspelmeyer2014}. With certain approximations, the coupling can be linearized to a beam-splitter or two mode squeezing interaction, which further enables the state transduction in a coherent manner \cite{zhong2022prapp}. Without loss of generality, we take a piezo-optomechanical-system-based transducer for demonstration, where the system's beam-splitter type Hamiltonian can be written in the rotating frame as \cite{han2020,zhong2020pra}
\begin{equation}
    \hat{H}/\hbar = g_\text{om}(\hat{a}^\dagger\hat{b}+\hat{a}\hat{b}^\dagger) + (g_\text{em}\hat{b}^\dagger\hat{c}+g^\ast_\text{em}\hat{b}\hat{c}^\dagger).
\end{equation}
We denote $\hat{a}$ and $\hat{c}$ as the bosonic operators for optical and electrical modes, respectively. $\hat{b}$ is an intermediate acoustic mode with frequency $\omega_\text{m}\sim\omega_\text{e}$. $g_\text{om}$ ($|g_\text{em}|$) is the optomechanical (piezo-mechanical) coupling strength. We keep $g_\text{em}$ as complex since the coupling phase would be important and the reason will become clear in the coming section. The conversion between the optical and microwave photon can be understood as mode swaps enabled by the two beam-splitter interactions. In experiment, the optical (electrical) mode is further coupled to an optical fiber (a microwave transmission line) which carries the input and output signals with coupling rate $\kappa_\text{o,c}$ and $\kappa_\text{e,c}$, respectively. In practice, all modes also intrinsically couple to their bath with the corresponding rates given by $\kappa_\text{o,i}$, $\kappa_\text{e,i}$ and $\kappa_\text{m}$.

To quantify the optical-microwave transduction, we first write down the Heisenberg-Langevin equations for each mode and the corresponding input-output formula as
\begin{equation}\label{dyeq}      \dot{\textbf{a}}=\textbf{A}\textbf{a}+\textbf{B}\textbf{a}_\text{in},
        \textbf{a}_\textbf{out}=\textbf{B}^\text{T}\textbf{a}-\textbf{a}_\text{in},
\end{equation}
where we label the vectors $\textbf{a}=\{\hat{a},\hat{c},\hat{b}\}^\text{T}$, $\textbf{a}_\text{in}=\{\hat{a}_\text{in,c},\hat{a}_\text{in,i},\hat{c}_\text{in,c},\hat{c}_\text{in,i},\hat{b}_\text{in}\}^\text{T}$ and $\textbf{a}_\text{out}=\{\hat{a}_\text{out,c},\hat{a}_\text{out,i},\hat{c}_\text{out,c},\hat{c}_\text{out,i},\hat{b}_\text{out}\}^\text{T}$. The lower indexes ``in/out" are used to indicate the input and output traveling modes, while ``c/i" represents the coupling and intrinsic loss ports. The matrices 
\begin{equation}
    \textbf{A}=
    \begin{pmatrix}
    -\frac{\kappa_\text{o}}{2}& 0 & -ig_\text{om}\\
    0 & -\frac{\kappa_\text{e}}{2} & -ig^\ast_\text{em}\\
    -ig_\text{om} & -ig_\text{em} & -\frac{\kappa_\text{m}}{2} 
    \end{pmatrix}
\end{equation}
and
\begin{equation}
    \textbf{B}=
    \begin{pmatrix}
    \sqrt{\kappa_\text{o,c}} &\sqrt{\kappa_\text{o,i}} & 0 & 0 & 0\\
    0 & 0 & \sqrt{\kappa_\text{e,c}} & \sqrt{\kappa_\text{e,i}} & 0 \\
    0 &0 &0& 0& \sqrt{\kappa_\text{m}}
    \end{pmatrix}.
\end{equation}
The Eq.~\ref{dyeq} is usually solved in the frequency domain, where one can obtain the connection between the input and output modes as the scattering relation
\begin{equation}\label{tlchannel}
    \textbf{a}_\text{out}[\omega]=\textbf{S}[\omega]\cdot\textbf{a}_\text{in}[\omega]
\end{equation}
where $\textbf{S}[\omega]=\textbf{B}^\text{T}(-i\omega \textbf{I}_3-\textbf{A})^{-1}\textbf{B}-\textbf{I}_5$ is the scattering matrix. Based on the matrix, we can identify the quantum transduction channel in both directions, e.g., in the narrow bandwidth limit ($\omega=0$), the microwave-to-optical conversion channel can be written down as
\begin{equation}
    \hat{a}_\text{out,c}=\sqrt{\eta}\hat{c}_\text{in,c}+\sqrt{1-\eta}\hat{e}.
\end{equation}
It is a beam splitter interaction mixing the microwave signal and noisy inputs. $\eta$ is the transmissivity which can be expressed as
    $\eta={4C_\text{om}C_\text{em}\zeta_\text{o}\zeta_\text{e}}/{(1+C_\text{om}+C_\text{em})^2}$, where $\zeta_\text{o}=\kappa_\text{o,c}/\kappa_\text{o}$ and $\zeta_\text{e}=\kappa_\text{e,c}/\kappa_\text{e}$ are the extraction ratios, and $C_\text{om}=4g_\text{om}^2/\kappa_\text{o}\kappa_\text{m}$ and $C_\text{em}=4|g_\text{em}|^2/\kappa_\text{e}\kappa_\text{m}$ are the system cooperativities. $\hat{e}$ collects the noisy influences from all other input modes
\begin{equation}
    \hat{e}=\frac{1}{\sqrt{1-\eta}}(S_{11}\hat{a}_\text{in,c}+S_{12}\hat{a}_\text{in,i}+S_{14}\hat{c}_\text{in,i}+S_{15}\hat{b}_\text{in}),
\end{equation}
where $S_\text{ij}$ are the corresponding scattering matrix elements. Obviously, the microwave-optical conversion defines a bosonic thermal loss channel with transmissivity $\eta$ and thermal noise $n_e=\braket{\hat{e}^\dagger\hat{e}}$. We denote it as $\mathcal{N}(\eta,n_e)$. It is known that this channel is a Gaussian channel, which maps an input Gaussian state with covariance matrix $\mathbf{V}$ into $\mathbf{TVT}^\mathrm{T}+\mathbf{N}$, where $\mathbf{T}=\sqrt{\eta}\mathbf{I}_2$ and $\mathbf{N}=(1-\eta)(2n_e+1)\mathbf{I}_2$ \cite{weedbrook2012}. 

A quantum channel is capable of transmitting quantum information, and the transmission rate is quantified by quantum capacity. For many quantum channels, including the thermal loss channel, determining the exact quantum capacity is hard. As a result, lower bounds—being more tractable—are often used to analyze and understand channel properties. As mentioned, the quantum transducer is modeled as a bosonic thermal loss channel $\mathcal{N}(\eta,n_e)$, and a commonly used lower bound on the quantum capacity of such a channel is given by \cite{weedbrook2012}
\begin{equation}\label{clower}
    Q_\text{LB}^\mathcal{N}=\max\{0,\log_2\frac{\eta}{1-\eta}-g(n_e)\},
\end{equation}
where $g(x)\equiv(x+1)\log_2(x+1)-x\log_2x$ and it is a monotonically increasing function. This bound is tight for so called pure loss channel ($n_e=0$). For pure loss channel, it is easy to get that $\eta=1/2$ is the threshold to have a positive channel capacity. For general thermal loss channel, it is necessary to have $\eta>1/2$ in order to get a positive quantum capacity. The negative sign in front of $g(n_e)$ highlights the detrimental effect of thermal noise: as $n_e$ increases, the channel's quantum capacity decreases. Therefore, identifying effective strategies for suppressing thermal noise is of critical importance.

\begin{figure}[t]
\centering
\includegraphics[width=\columnwidth]{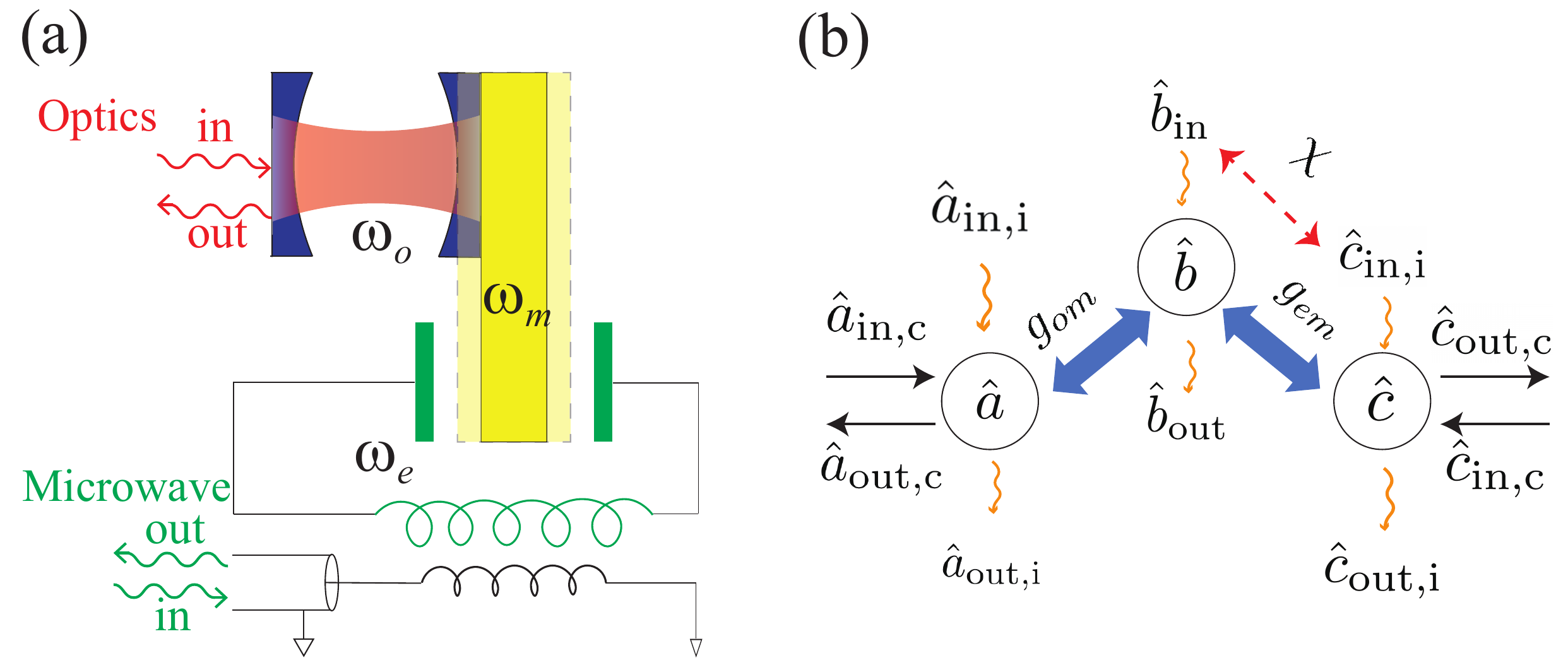}
\caption{(a) Schematic figure for piezo-optomechanical system used for direct quantum transduction. (b) Illustration for the input–output relations with correlated intrinsic dissipation between the microwave and acoustic modes.     \label{fig0}}
\end{figure}

\begin{figure*}[t]
\centering
\includegraphics[width=\textwidth]{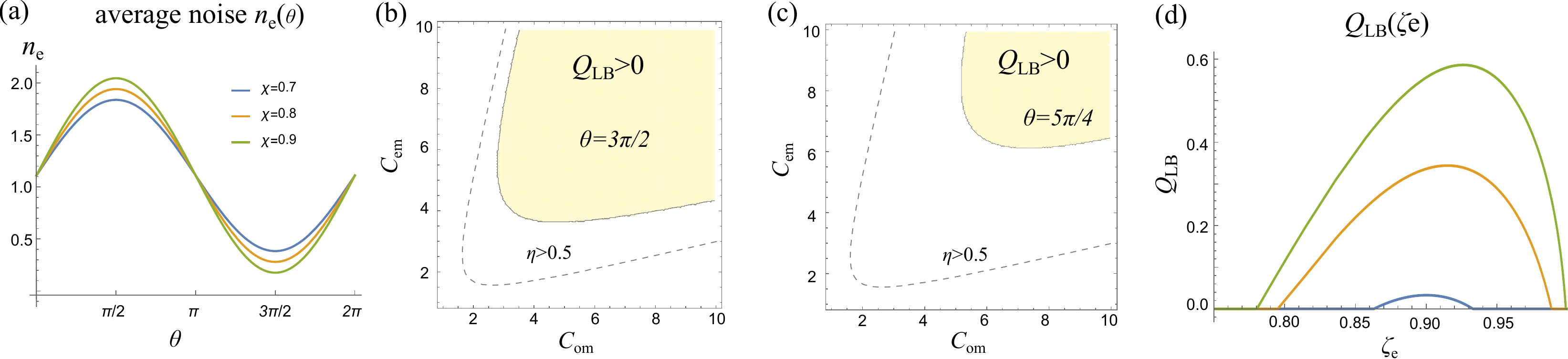}
\caption{(a) The average noise coupled to the transduction channel with respect to the piezo-mechanical coupling phase $\theta$. The three curves correspond to varied degrees of noise correlations $\chi=\{0.7,0.8,0.9\}$. $C_\text{om}=C_\text{om}=4$ and $\zeta_o=\zeta_e=0,9$ are used for all three curves. (b) and (c) show the channel's quantum capacity lower bound by scanning the system parameters $C_\text{om}$ and $C_\text{em}$ with $\theta=5\pi/4$ and $\theta=3\pi/2$, respectively. The degree of noise correlation $\chi=0.9$ is used for both (b) and (d). The dashed lines delineate the regime where the channel's transmissivity $\eta>0.5$ (the threshold). (d) The quantum capacity lower bound in terms of the electrical mode extraction ratio. The blue, orange and green lines differ in the parameters $\{C_\text{om},C_\text{em}\}=\{4,4\},\{5,5\} \text{ and } \{6,6\}$, respectively. For demonstration, the optical mode extraction ration $\zeta_o=0.9$ and the bath temperature $T=1$ K are used for all plots.      \label{fig1}}
\end{figure*}

\textit{Independent noise model}---Typically, the noise terms originate from different physical process and are thus independent. For the piezo-optomechanical transducer, all modes are intrinsically coupled to their thermal environment. Assuming the temperature is ${T}$, the average noise can be estimated by $n_\text{th}^\text{o,m,e}\equiv(e^{{\hbar\omega_\text{o,m,e}/k_\text{B}{T}}}-1)^{-1}$. The optical mode is well known for its insensitivity to the environment since it has a relatively large mode frequency (several hundreds of THz) and $n^\text{o}_\text{th}=\braket{\hat{a}_{in}^\dagger\hat{a}_{in}}$ is very small even at room temperature. For temperature in the range of mK, the electrical and acoustic modes with much lower frequencies (generally $\omega_\text{e,m}$ are around several GHz) have a non-negligible thermal noises. If we ignore safely the optical noise, the average input noise photon is written down as
\begin{equation}\label{eqnoise}
 n_e=\braket{\hat{e}^\dagger\hat{e}}=\frac{1}{1-\eta}(|S_{14}|^2n^\text{e}_\text{th}+|S_{15}|^2n^\text{m}_\text{th}),
\end{equation}
where $|S_{14}|^2={4C_\text{em}C_\text{om}}\zeta_\text{o}(1-\zeta_\text{e})/{(1+C_\text{om}+C_\text{em})^2}$ and $|S_{15}|^2={4C_\text{om}}\zeta_\text{o}/{(1+C_\text{om}+C_\text{em})^2}$. $n^\text{e}_\text{th}=\braket{\hat{c}^\dagger_\text{in,i}\hat{c}_\text{in,i}}$ and $n^\text{m}_\text{th}=\braket{\hat{b}^\dagger_\text{in}\hat{b}_\text{in}}$ are the electrical and acoustic thermal noise, respectively. In experimental design \cite{han2020,meesala2024prx}, the acoustic and microwave modes are coupled to the same bath, thus we denote $n^\text{e}_\text{th}=n^\text{m}_\text{th}\equiv n_\text{th}$, which simplifies the input noise expression as
\begin{equation}
n_e=\frac{1}{1-\eta}(|S_{14}|^2+|S_{15}|^2)n_\text{th}.
\end{equation}

\textit{Correlated noise model}---As mentioned earlier, the noise $n_\text{e}$ in the transduction channel will degrade its performance. In this section, we show this noise can be reduced if $\hat{b}_\text{in}$ and $\hat{c}_\text{in,i}$ are correlated. Importantly, we emphasize that this noise reduction is not limited to a specific platform---it exists in principle for any system with correlated noise inputs. Therefore, the protocol we propose below is broadly applicable and of potential interest across a wide range of quantum technologies. In this letter, we take the practical setting of piezo-optomechanical transducer for example, where the electrical and acoustic modes are intrinsically coupled to the same bath, which is locally affected by the same laser heating. Explicitly, we assume that their correlation formula can be written down as
\begin{equation}
    \braket{\hat{c}_\text{in,i}^\dagger\hat{b}_\text{in}}\equiv n_\text{th}\cdot\chi,
\end{equation}
where $\chi$ is usually a complex number satisfying $0 \le|\chi|\le 1$, and it can be understood as a measure of the degree of correlations. Without losing any physical meaning, we take $\chi$ as a real positive number in the following discussions. As a result, the average input noise in Eq.~\ref{eqnoise} is modified as
\begin{equation}
    n_e=\frac{1}{1-\eta}(|S_{14}|^2+|S_{15}|^2+S_{14}^\ast S_{15}\chi+S_{15}^\ast S_{14}\chi)n_\text{th}.
\end{equation}
We see there are two extra terms added in the average noise, and the summation of them is obviously real. If it is negative in the mean time, the average noise $n_e$ can be reduced. Interestingly, it is shown that the sign of the summation is controlled by the piezo-mechanical coupling phase. This can be understood intuitively---the correlated noise is injected into the acoustic and the electrical modes, so the phase of their coupling becomes important since it determines the way the correlated noise interact in a coherent manner. Specifically, we write out the piezo-mechanical coupling as $g_\text{em}=|g_\text{em}|e^{i\theta}$ with $\theta$ the coupling phase. The scattering elements $S_{14}$ and $S_{15}$ take the form
\begin{equation}
    S_{14}=\frac{-2e^{i\theta}\sqrt{C_\text{em}C_\text{om}(1-\zeta_e)\zeta_o}}{1+C_\text{om}+C_\text{em}}, S_{15}=\frac{-2i\sqrt{C_\text{om}\zeta_o}}{1+C_\text{om}+C_\text{em}}.
\end{equation}
Thus, we have
\begin{equation}\label{eq14}
    S_{14}^\ast S_{15}+S_{15}^\ast S_{14}=\frac{8\sqrt{C_\text{em}(1-\zeta_e)}C_\text{om}\zeta_o}{(1+C_\text{om}+C_\text{em})^2}\sin\theta,
\end{equation}
which is negative if the phase satisfies $\pi<\theta<2\pi$, thus contributing to the noise suppression. It is worth mentioning when $\theta\in(0,\pi)$, the average noise will increase due to the noise correlation. Specially, when $\theta=0$ or $\pi$, the noise correlation does not contribute to the overall noise, which is effectively the same as the independent noise model. 

\textit{Numerical results}---Assuming the transducer is in $T=1$K environment, the thermal noise $n_\text{th}\simeq 1.6$ for the modes with frequency $\omega_\text{m,e}=10$ GHz. We first calculate the average noise in terms of the coupling phase. As shown in Fig.~\ref{fig1}(a), for different degrees of noise correlation $\chi$, the average noise $n_e(\theta)$ can almost be suppressed to below one half, which is critical for achieving single-photon-level quantum transduction. This technique has the potential to significantly reduce the reliance on cryogenic cooling, alleviating one of the major hardware constraints in quantum device operation. 

Figure~\ref{fig1}(b) and \ref{fig1}(c) depict the quantum capacity lower bound of the transduction channel according to Eq.~\ref{clower}. In Fig.~\ref{fig1}(b), the yellow region on the upper right corner shows the parameter regime of positive capacity lower bound, where the system cooperativities $C_\text{om}$ and $C_\text{em}$ are scanned. The dashed black line encloses the upper-right corner where the system parameters fulfill $\eta>1/2$---a potential regime for $Q_\text{LB}>0$ depending on the channel noise. We see the yellow area almost occupy all the enclosure for $\theta=3\pi/2$, indicating a significant noise suppression. For slightly smaller coupling phase $\theta=5\pi/4$, the yellow area shrinks quickly since the average noise increases, matching the results in Fig.~\ref{fig1}(a). In fact, if we keep decreasing the coupling phase, e.g., $\theta=\pi$, the yellow blob disappears totally in the same scanned parameter regime, indicating a zero transduction capacity without the assistance from noise correlations. 

Due to the noise correlation, the transduction capacity exhibits a non-trivial dependence on the electrical extraction ratio. On the one hand, the channel transmissivity $\eta$ is proportional to $\zeta_e$, which means a larger $\zeta_e$ generally benefits quantum transduction. However, according to Eq.~\ref{eq14}, increasing $\zeta_e$ will lead to a smaller contribution from the correlated noise, thereby diminishing the potential gain in quantum capacity. This aligns with the intuition that a large $\zeta_e$ means less influence from the intrinsic noise $\hat{c}_\text{in,i}$, while this intrinsic noise is exactly what we need to cancel part of the noise from $\hat{b}_\text{in}$. In Fig.~\ref{fig1}(d), we show this trade off for varied system cooperativities. The bule, orange and green curves represent the capacity lower bound as $\zeta_e$ is tuned. All of them rise in the beginning and quickly go down as we keep tuning up $\zeta_e$, with a peak around $\zeta_e=0.9$. In experiments, optimizing $\zeta_e$ is critical to achieving the best transduction performance.

\textit{Discussion}---Correlated noise is present in many practical systems and has been investigated across a variety of platforms for different applications \cite{buscemi2010,clemens2004,harper2023,novais2006,salhov2024}. For instance, it has been widely recognized that quantum error correction schemes based on independent noise models are often insufficient in realistic scenarios, prompting the development of error correction techniques that explicitly incorporate correlated noise \cite{de2005,novais2006,harper2023}. In this work, we developed and demonstrated a strategy to enhance the performance of a quantum transducer by harnessing noise correlations. The noise correlation with proper control is shown to suppress the overall noise to sub-photonic level even the environment temperature is set to $1$ K, effectively achieving system cooling.

When combined with other cooling methods, such as radiative cooling \cite{mingrui2020,zhong2020prl}, this technique has the potential to achieve even greater noise suppression, thereby further improving quantum transduction performance. It is important to note, however, that noise correlations can also amplify the overall noise under certain conditions. Specifically, when the coupling phase $\theta\in(0,\pi)$ and the noise correlation coefficient $\chi$ remains positive, it would lead to increased overall noise. In contrast, if $\chi<0$, the noise can be suppressed within the same phase range. Therefore, in pratical transducer design, the coupling phase should be carefully chosen based on the specific characteristics of the noise environment.

{It is also worth noting that noise correlations may not remain stable over time due to fluctuations in the environment or system parameters. A key advantage of the proposed protocol is its robustness---even in the absence of beneficial noise correlations, it performs at least as well as conventional approaches that do not exploit correlation. This ensures that the protocol does not introduce additional vulnerability and remains effective across a range of practical conditions.}

Beyond direct applications in quantum transduction, we believe that the exploration of noise correlations and the protocol proposed in this work could benefit a much broader range of quantum technologies. For example, it may prove useful in enhancing entanglement generation in parametrically coupled systems, where the noise is typically an issue of degrading the entanglement fidelity \cite{meesala2023np,rueda2019,Barzanjeh2011}. We leave further exploration of these possibilities to future work.

\begin{acknowledgments}
C.Z. thanks the start-up support from Xi'an Jiaotong University (Grant No. 11301224010715).
\end{acknowledgments}

\bibliography{all}

\onecolumngrid




\end{document}